\documentclass{PoS}

\usepackage{siunitx} 
\usepackage{csquotes}
\usepackage{amsmath}
\usepackage{hepnames}
\DeclareRobustCommand{\PQ}{\HepGenParticle{Q}{}{}\xspace} 
\DeclareRobustCommand{\PaQ}{\HepGenAntiParticle{Q}{}{}\xspace} 
\usepackage{sidecap}

\title{NLO Corrections to Heavy Flavour Distributions in Polarized Deep-Inelastic Scattering}

\ShortTitle{NLO HQ distributions in polarized DIS}

\author{\speaker{F. Hekhorn} and M. Stratmann\\
        Institute for Theoretical Physics, University of T\"ubingen, 
        Auf der Morgenstelle 14,\\ 72076 T\"ubingen, Germany\\
        E-mail: \email{felix.hekhorn@uni-tuebingen.de}, \email{marco.stratmann@uni-tuebingen.de}}

\abstract{We present a first calculation of the heavy flavor contribution 
to the longitudinally polarized DIS structure function $g_1$, 
differential in the transverse momentum or the rapidity of the observed heavy antiquark 
$\PaQ$. All results are obtained at next-to-leading order accuracy with a 
newly developed parton-level Monte Carlo generator that also allows one 
to study observables associated with the heavy quark pair such as its invariant mass 
distribution or its correlation in azimuthal angle. First phenomenological studies 
are presented in a kinematic regime relevant for a future Electron-Ion Collider 
with a particular emphasis on the sensitivity to the helicity gluon distribution. 
Finally, we also provide first NLO results for the full neutral-current sector of 
polarized DIS, i.e., including contributions from $\PZ$-boson exchange.}

\FullConference{XXVII International Workshop on 
Deep-Inelastic Scattering and Related Subjects - DIS2019\\
		8-12 April, 2019\\
		Torino, Italy}

\begin{document}
In this talk, we present for the first time two important extensions to our previous
work on inclusive heavy quark (HQ) polarized electroproduction 
at next-to-leading order accuracy \cite{Hekhorn:2018ywm}.
Firstly, we discuss various exclusive distributions related to the HQ DIS structure function
$g_1^Q$ differential in the transverse momentum or rapidity of the HQ or in
the invariant mass and the correlation in azimuthal angle of the HQ pair. These results
have been obtained with a newly developed parton-level Monte Carlo generator.
Secondly, we provide the electroweak contributions to the neutral current (NC) sector in DIS.

In our first paper \cite{Hekhorn:2018ywm}, we followed the concept of the phase-space 
slicing method \cite{Laenen:1992zk}, dealing with all phase-space integrals 
in a largely analytic way. This restricts one to the computation of single-inclusive distributions
either in the rapidity $y$ or in the transverse momentum $p_T$ of the HQ.
To set up a general parton-level Monte Carlo generator for HQ production in polarized DIS, 
that evaluates all phase-space integrations numerically, we adopt the subtraction method 
based on generalized plus distributions \cite{Harris:1995tu,Mangano:1991jk} to isolate all 
singular regions in phase-space.
With the new program at hand, we get access to basically any HQ distribution of interest
including correlated distributions in the invariant mass or the azimuthal angle
of the HQ pair.
We leave the discussion of the relevant technical details 
to forthcoming publications \cite{paper2,PhD} and focus here 
on the most important phenomenological results.

\begin{SCfigure}
\includegraphics[width=.5\textwidth]{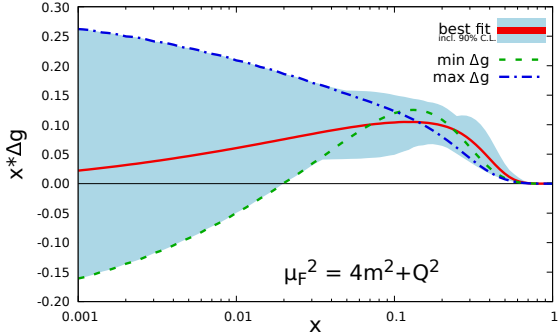}
\caption{Polarized gluon distribution $\Delta\Pg(x,\mu_F^2)$ 
of the DSSV group \cite{deFlorian:2014yva} at the scale $\mu_F^2 = \SI{19}{\GeV}$
relevant for charm electroproduction,
including the best fit (solid line) and two extreme variations 
("max $\Delta g$" and "min $\Delta g$") allowed by present uncertainties (shaded band).}
\label{fig:xg}
\end{SCfigure}
The main asset of HQ production is its dominance of gluon-induced processes, in case
of DIS photon-gluon fusion, already at the lowest order (LO) approximation of QCD.
To estimate how the measurement of HQ production in polarized DIS at a future
Electron-Ion Collider (EIC) \cite{Boer:2011fh}
can help to further our knowledge of the so far 
poorly constrained gluon helicity density $\Delta g$, we show in Fig.~\ref{fig:xg}
the current uncertainties which are particularly pronounced for small momentum
fractions $x$, say below $x\simeq 0.01$. Apart from the best fit result given by
the DSSV group \cite{deFlorian:2014yva}, we will also utilize in what follows 
their uncertainty estimates, in particular, the two extreme sets labeled
as "max $\Delta g$" and "min $\Delta g$". For our calculations, we use
the MSTW NLO set of unpolarized PDFs \cite{Martin:2009iq} which was also
adopted in the DSSV analysis to ensure constraints from positivity.

\begin{figure}[ht]
\begin{center}
\includegraphics[width=.4\textwidth]{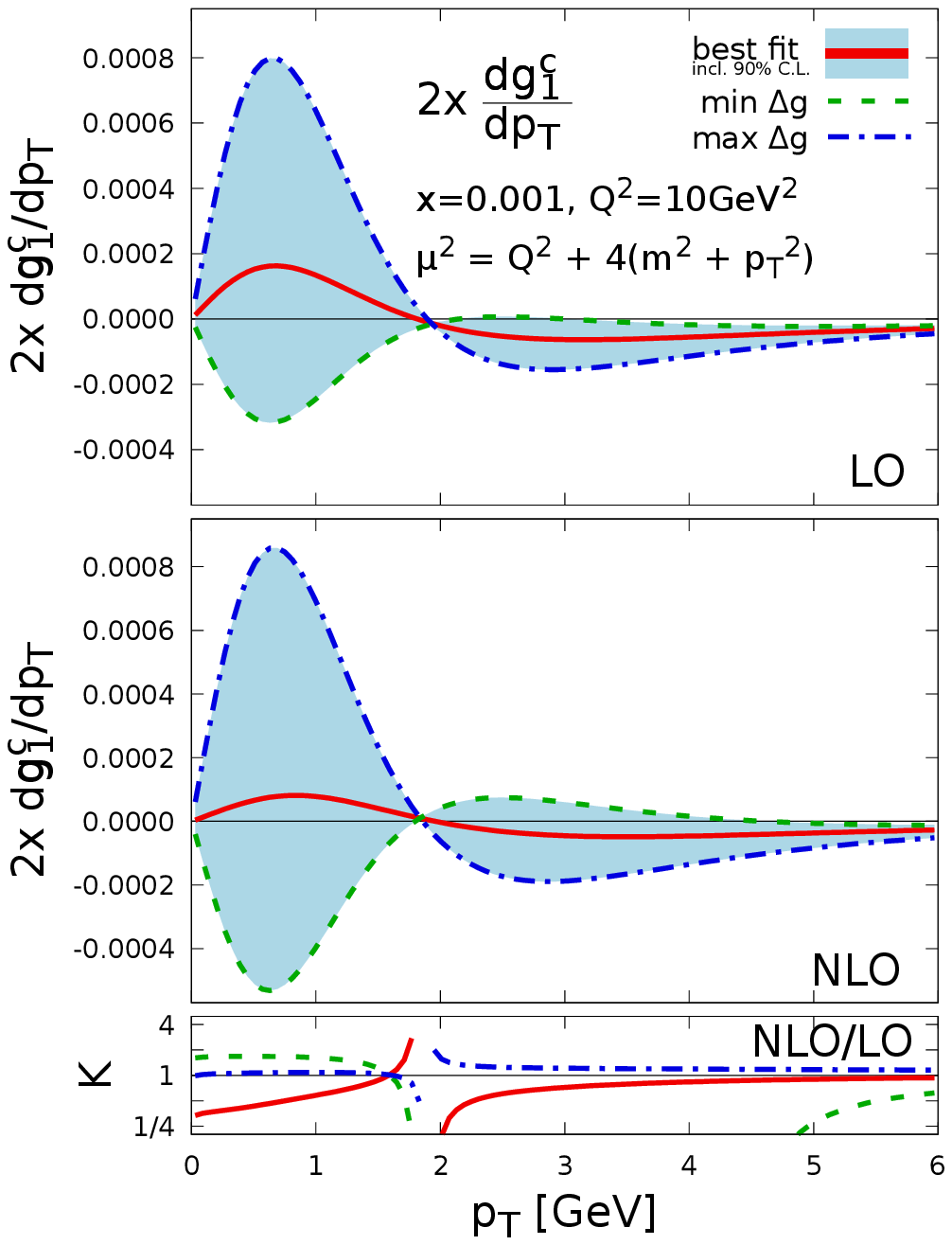}
\includegraphics[width=.4\textwidth]{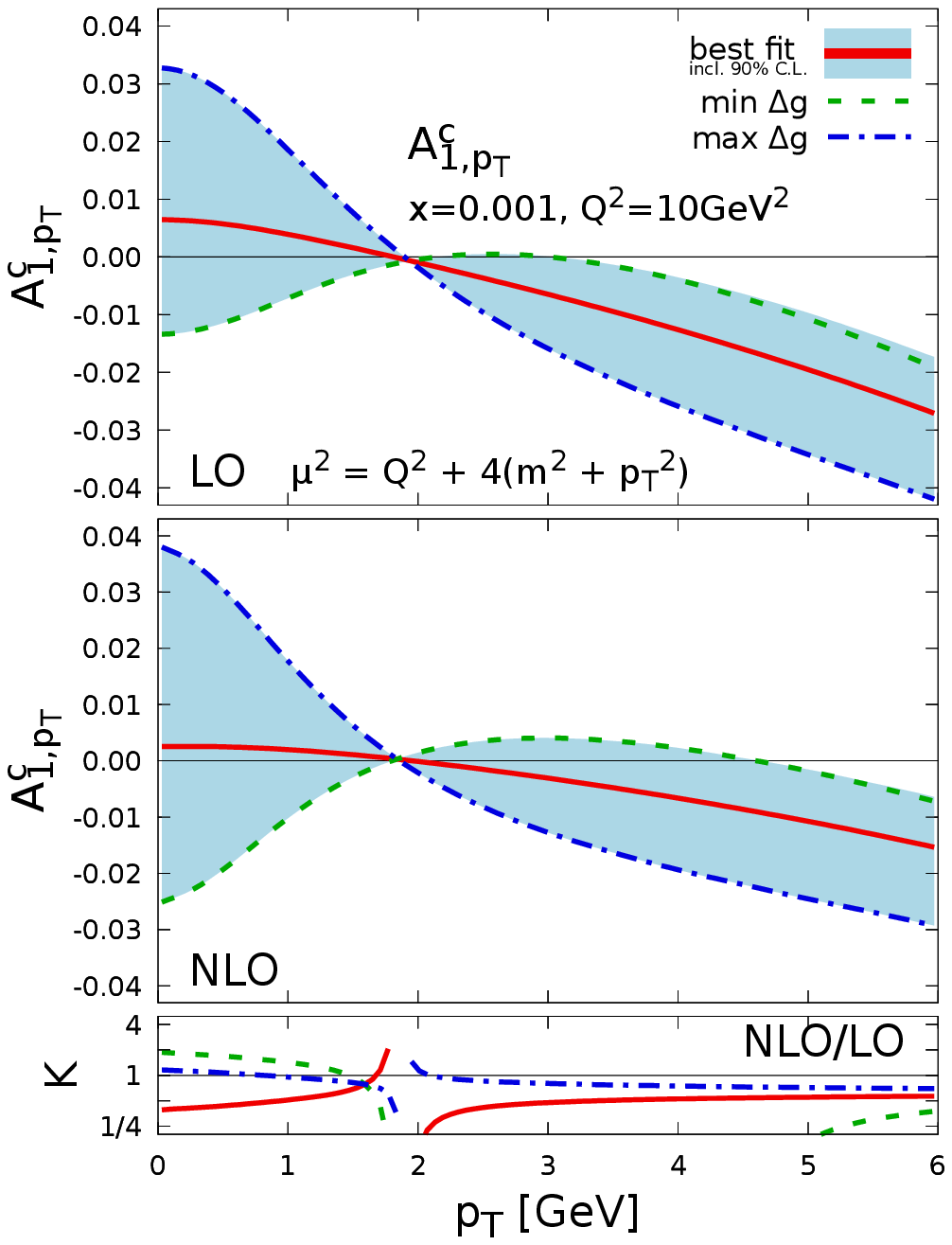}
\end{center}
\vspace{-0.65cm}
\caption{LO (top) and NLO (middle panel) results for the transverse momentum distribution
$2x dg_1^{\Pqc}/dp_T$ (left) and the associated double-spin asymmetry
$A_{1,p_T}^{\Pqc}$ (right) as a function of $p_T$ for fixed $x=10^{-3}$ and $Q^2=10\,\mathrm{GeV}^2$.
The different lines and the shaded band correspond to
different choices of helicity gluon PDFs as shown in Fig.~\ref{fig:xg}.
All results are obtained for our default choice of scale $\mu_0^2 = Q^2 + 4(m^2+p_T^2)$
with a charm quark mass of $m=1.5\,\mathrm{GeV}$.
The lower panels show the ratios of NLO and LO results (''$K$-factor'').}
\label{fig:dpt}
\end{figure}
As a first example, we investigate the single-inclusive
$p_T$-distribution $dg_1^{\Pqc}/dp_T$, for which we define also 
a corresponding $p_T$-dependent double-spin asymmetry
\begin{align}
A_{1,p_T}^{\Pqc}(x,Q^2,p_T) &= \frac{dg_1^{\Pqc}/dp_T}{dF_1^{\Pqc}/dp_T}\,. \label{eq:A1pt}
\end{align}
Both are shown in Fig.~\ref{fig:dpt} for fixed momentum fraction $x=10^{-3}$ and
photon virtuality $Q^2=10\,\mathrm{GeV}^2$ in a range of $p_T$ which should be
accessible at a future EIC. 
As we consider HQs with mass $m$, we obtain finite results even for $p_T=0$.
The asymmetry $A_{1,p_T}^{\Pqc}$ reaches measurable, percent-level values and exhibits a
strong sensitivity to $\Delta g$, clearly dependent on its size and sign as can be gathered
from the results obtained for the minimal and maximal $\Delta\Pg$.
\begin{SCfigure}
\includegraphics[width=.38\textwidth]{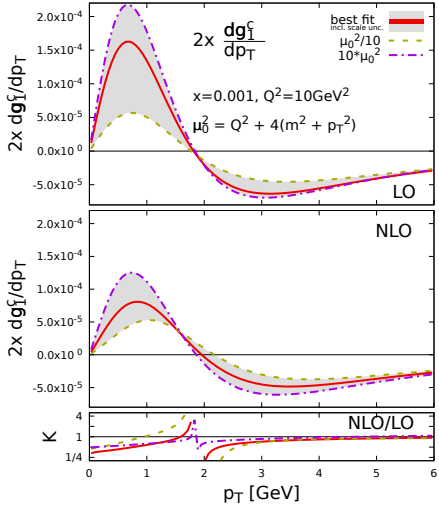}
\caption{Scale dependence of $2x\, dg_1^{\Pqc}/dp_T$ shown in
Fig.~\ref{fig:dpt} at LO (top) and NLO (middle) accuracy
for simultaneous variations of $\mu_R^2$ and $\mu_F^2$
by a factor of 10 around the default choice $\mu_0^2 = Q^2+4(m^2+p_T^2)$.
The bottom panel shows the variation of the $K$-factor.}
\label{fig:dpt-mu}
\end{SCfigure}

\begin{figure}[ht]
\begin{center}
\includegraphics[width=.45\textwidth]{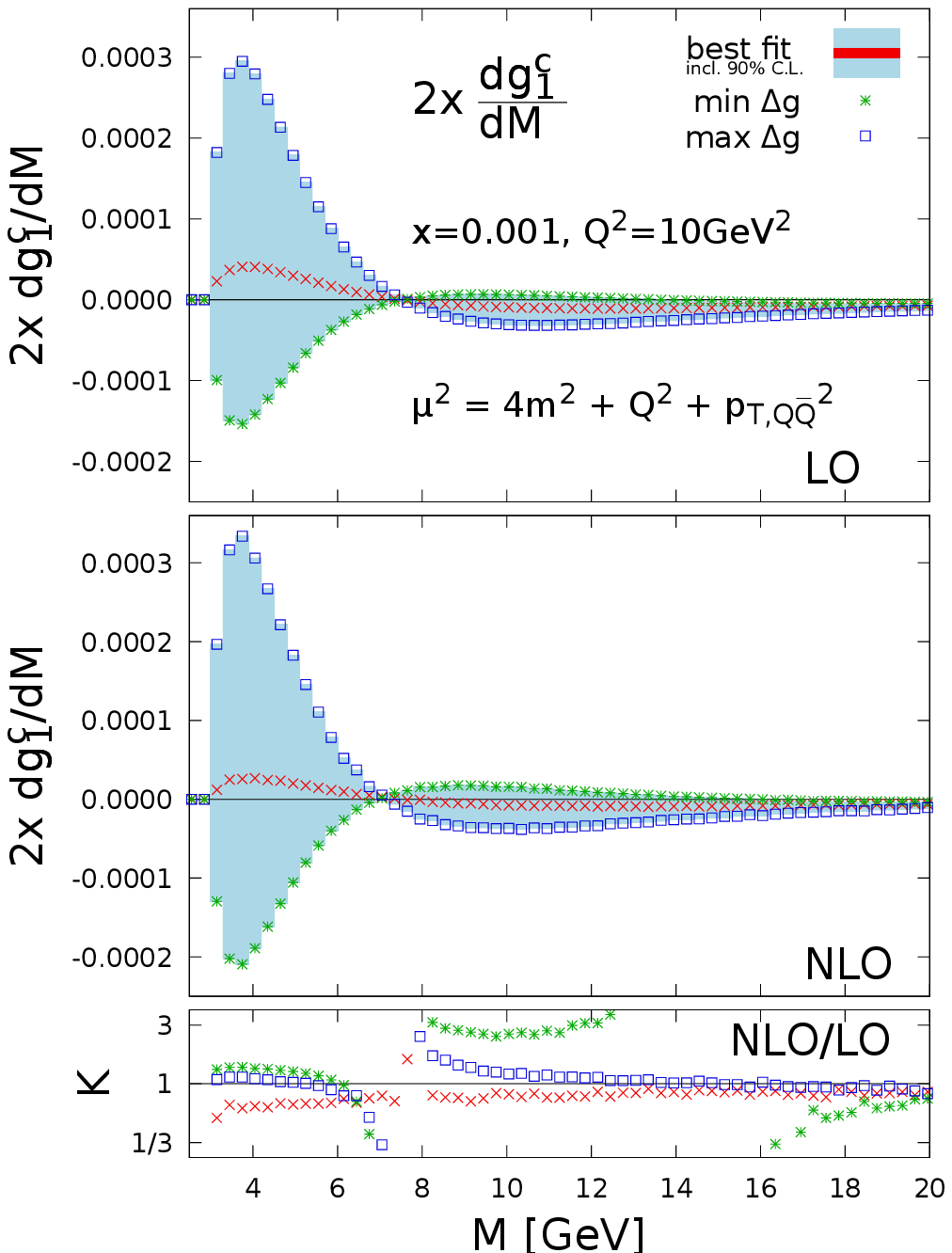}
\includegraphics[width=.45\textwidth]{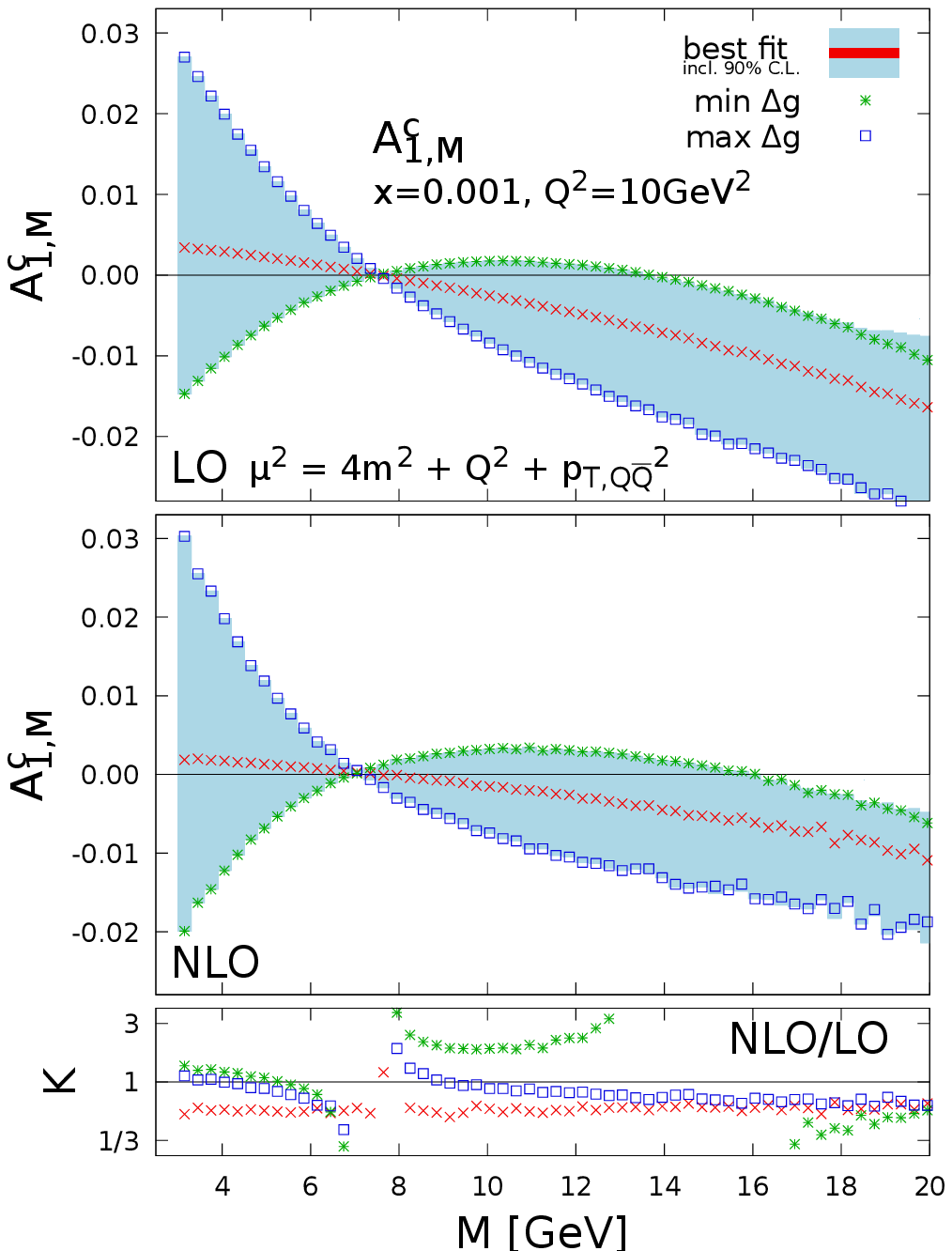}
\end{center}
\vspace{-1.0em}
\caption{As in Fig.~\ref{fig:dpt} but now as a function of the invariant mass $M$ of the
HQ pair: $2x dg_1^{\Pqc}/dM$ (left panels), $A_{1,M}^{\Pqc}$ (right panels).
Here, the default choice of scale is $\mu^2 = Q^2 + 4m^2 + p_{T,\PQ\PaQ}^2$, see text.}
\label{fig:dM}
\end{figure}
In Fig.~\ref{fig:dpt-mu} we illustrate the theoretical uncertainties associated with 
simultaneous variations of the renormalization ($\mu_R$) and factorization ($\mu_F$) 
scales for the $p_T$-distribution $2x\, dg_1^{\Pqc}/dp_T$ computed with the best fit of the DSSV set.
As is expected, the scale dependence is significantly reduced at NLO accuracy and,
more importantly, its size is much smaller than the spread in the results due to
current uncertainty in $\Delta g$ as shown in Fig.~\ref{fig:dpt}.
We note that including a $p_T$-dependence in the choice of scale $\mu$ is important to
obtain sensible results in the high-$p_T$ region.
 
As a first example that can be only obtained with our newly developed Monte-Carlo code, 
we study in Fig.~\ref{fig:dM} the invariant mass distribution of the HQ pair, i.e.,
the differential polarized DIS structure function $2x dg_1^{\Pqc}/dM$ 
and its associated double-spin asymmetry $A_{1,M}^{\Pqc}$. 
The latter is defined in an analogous way to Eq.~\eqref{eq:A1pt}.
Again, we choose $x=10^{-3}$ and $Q^2=10\,\mathrm{GeV}^2$ to fix the DIS kinematics in the
EIC range. Here, we set the default scale to $\mu^2 = Q^2 + 4m^2 + p_{T,\PQ\PaQ}^2$ as in
Ref.~\cite{Harris:1995tu} with $p_{T,\PQ\PaQ}$ referring to the 
transverse momentum of the HQ pair. As for the single-inclusive $p_T$ distributions
shown in Fig.~\ref{fig:dpt}, the double-spin asymmetry exhibits a nice sensitivity the
the sign and size of the helicity gluon density adopted in the calculation. Again,
the expected values for $A_{1,M}^{\Pqc}$ are in the percent level range which should
be accessible with the envisioned high luminosity of a future EIC.

Next, we turn to our recent computation of the electroweak contributions to the neutral current (NC) 
DIS structure functions at NLO accuracy \cite{PhD,felix} 
which are mediated by the exchange of a $\PZ$ boson.
Here, we will consider bottom quark production to optimize the kinematical conditions
and use the PDF sets of the NNPDF group \cite{Ball:2012cx} throughout.
In order to establish a common notation we denote the coupling of a boson $b$ to a 
fermion $f$ as $-i e \Gamma^\mu_{b,f}$ with $e$ the universal electric charge and
\begin{align}
\Gamma^\mu_{b,f} &= g^V_{b,f} \Gamma^\mu_{V} + g^A_{b,f} \Gamma^\mu_{A} = g^V_{b,f} \gamma^\mu + g^A_{b,f} \gamma^\mu\gamma^5, \quad b\in\{\Pgg,\PZ\},f\in\{\Pl,\Pq,\PQ\}\;. \label{eq:couplingPhZ}
\end{align}
To deal with the subtleties of $\gamma_5$ in higher order calculations
we adopt the MVV scheme \cite{Moch:2015usa}.

\begin{figure}[ht]
\begin{center}
\includegraphics[width=.8\textwidth]{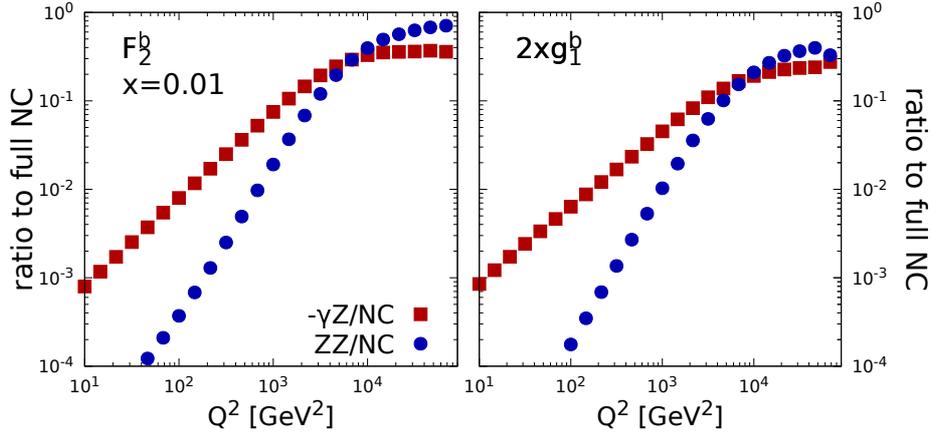}
\end{center}
\vspace{-2.0em}
\caption{Relative contributions of $R^{\Pgg\PZ}_{H}$ (squares) and 
$R^{\PZ\PZ}_{H}$ (circles) to the full NC parity-conserving structure functions 
$F_2^{\Pqb}$ (left) and $2xg_1^{\Pqb}$ (right) as a function of $Q^2$ for
fixed $x=0.01$. 
Note that the $\Pgg\PZ$-interference is plotted with a relative minus sign.}
\label{fig:R}
\end{figure}
We define the relative contributions of the $\gamma Z$ and $Z$ boson exchange 
to the full NC, parity-conserving HQ DIS structure functions $H\in\{F_2^{\Pqb},2xg_1^{\Pqb}\}$ by
\begin{equation}
R^{\Pgg\PZ}_{H}(x,Q^2) = 
\frac{(g_{\PZ,\Pe}^V-g_{\PZ,\Pe}^A)   \eta_{\Pgg\PZ}   H^{\Pgg\PZ}(x,Q^2)}{H^{NC}(x,Q^2)}
\;\;\;\text{and}\;\;\;
R^{\PZ\PZ}_{H}(x,Q^2) = 
\frac{(g_{\PZ,\Pe}^V-g_{\PZ,\Pe}^A)^2 \eta_{\Pgg\PZ}^2 H^{\PZ\PZ}(x,Q^2)}{H^{NC}(x,Q^2)}\;,
\end{equation}
respectively. The factor $\eta_{\Pgg\PZ}$ accounts for the difference between the photon and 
the $\PZ$-propagator. The relevance of the electroweak contributions is 
illustrated in Fig.~\ref{fig:R}. One can deduce two important results. 
Firstly, the $\Pgg\PZ$-interference term and the 
$\PZ$-contribution do not exhibit the same $Q^2$-dependence, which can be traced back to fact
that we deal here with massive quarks. Secondly, as expected, the electroweak contributions
become increasingly relevant the higher the $Q^2$ and they should not be neglected in analyses
of DIS data once $Q^2 \gtrsim 10^{3}\,\mathrm{GeV^2}$.
 
\begin{figure}[ht]
\begin{center}
\includegraphics[width=.8\textwidth]{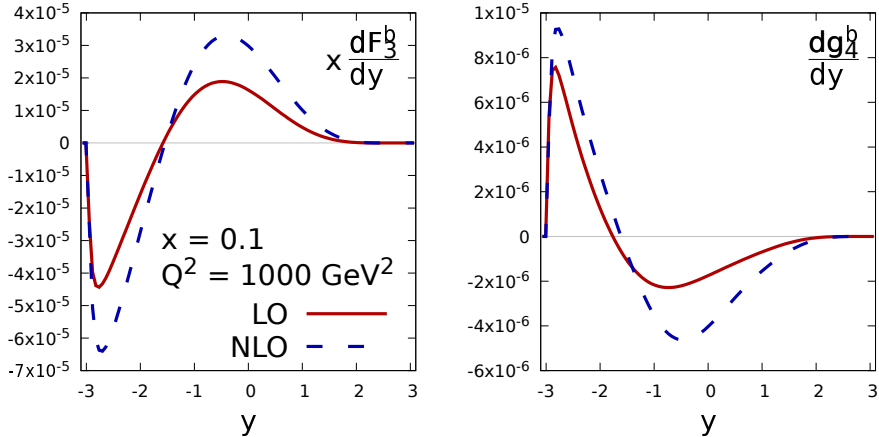}
\end{center}
\vspace{-2.0em}
\caption{Rapidity distribution of the parity-violating structure functions $xF_3^{\Pqb}$ (left) and $g_4^{\Pqb}$ (right) in LO (solid) and NLO (dashed) accuracy for $Q^2=10^3\,\mathrm{GeV}^2$
and $x=0.1$.}
\label{fig:SFPV}
\end{figure}
Finally, with the inclusion of the electroweak exchanges, we also get access to the 
parity-violating DIS structure functions \cite{Tanabashi:2018oca} 
$xF_3^{\Pqb}$, $g_4^{\Pqb}$, and $g_L^{\Pqc}$ 
which are given in Fig.~\ref{fig:SFPV}. 
As they do not contribute in LO to the fully inclusive DIS cross section and, in addition, 
vanish in the case of $p_T$-differential distributions \cite{PhD}, 
we shown in Fig.~\ref{fig:R} their rapidity dependence for $Q^2=10^3\,\mathrm{GeV}^2$
and $x=0.1$. The LO results were already discussed long ago in Ref.~\cite{Leveille:1978px} but
-- to the best of our knowledge -- the NLO results are presented here for the first time.


\begin{thebibliography}{99}
%
\bibitem{Hekhorn:2018ywm} 
  F.~Hekhorn and M.~Stratmann,
  Phys.\ Rev.\ D {\bf 98}, 014018 (2018);
%
  PoS DIS {\bf 2018}, 155 (2018).
%
\bibitem{Laenen:1992zk} 
  E.~Laenen, S.~Riemersma, J.~Smith, and W.~L.~van Neerven,
  Nucl.\ Phys.\ B {\bf 392}, 162 (1993).
%
\bibitem{Harris:1995tu} 
  B.~W.~Harris and J.~Smith,
  Nucl.\ Phys.\ B {\bf 452}, 109 (1995).
%
\bibitem{Mangano:1991jk} 
  M.~L.~Mangano, P.~Nason, and G.~Ridolfi,
  Nucl.\ Phys.\ B {\bf 373}, 295 (1992);
  S.~Frixione, M.~L.~Mangano, P.~Nason, and G.~Ridolfi,
  Nucl.\ Phys.\ B {\bf 412}, 225 (1994).
%
\bibitem{paper2} 
  F.~Hekhorn and M.~Stratmann, in preparation.
%
\bibitem{PhD} 
  F.~Hekhorn, PhD thesis, University of T\"{u}bingen, 2019, in preparation.
%
\bibitem{Boer:2011fh} 
  D.~Boer {\it et al.},
  {\tt arXiv:1108.1713};
  A.~Accardi {\it et al.},
  Eur.\ Phys.\ J.\ A {\bf 52}, 268 (2016).
%
\bibitem{deFlorian:2014yva} 
  D.~de Florian, R.~Sassot, M.~Stratmann, and W.~Vogelsang,
  Phys.\ Rev.\ Lett.\  {\bf 113}, 012001 (2014).
%
\bibitem{Martin:2009iq} 
  A.~D.~Martin, W.~J.~Stirling, R.~S.~Thorne, and G.~Watt,
  Eur.\ Phys.\ J.\ C {\bf 63}, 189 (2009).
%
\bibitem{felix} 
  F.~Hekhorn, in preparation.
%
\bibitem{Ball:2012cx} 
  R.~D.~Ball {\it et al.},
  Nucl.\ Phys.\ B {\bf 867}, 244 (2013);
  E.~R.~Nocera {\it et al.} [NNPDF Collaboration],
  Nucl.\ Phys.\ B {\bf 887}, 276 (2014).
%
\bibitem{Moch:2015usa} 
  S.~Moch, J.~A.~M.~Vermaseren and A.~Vogt,
  Phys.\ Lett.\ B {\bf 748}, 432 (2015); 
  S.~A.~Larin,
  Phys.\ Lett.\ B {\bf 303}, 113 (1993);
  J.~A.~M.~Vermaseren,
  math-ph/0010025.
%
\bibitem{Tanabashi:2018oca} 
  For the notational conventions regarding parity-violating DIS structure functions, see
  M.~Tanabashi {\it et al.} [Particle Data Group],
  Phys.\ Rev.\ D {\bf 98}, 030001 (2018).
\bibitem{Leveille:1978px}
  J.~P.~Leveille and T.~J.~Weiler,
  Nucl.\ Phys.\ B {\bf 147}, 147 (1979);
  W.~Vogelsang and A.~Weber,
  Nucl.\ Phys.\ B {\bf 362}, 3 (1991).
%

\end{thebibliography}
\end{document}